\documentclass[aps,twocolumn,showpacs]{revtex4}
\usepackage{graphicx}
\usepackage{epsfig,color}
\usepackage{bm}

\begin{document}

\setcounter{page}{1}
\newcommand{\re}[1]{(\ref{#1})}
\newcommand{\lab}[1]{\label{#1}}
\newcommand{\ci}[1]{\cite{#1}}
\renewcommand{\baselinestretch}{1.25}
\newcommand{\bfr}{\begin{flushright}}
\newcommand{\bfl}{\begin{flushleft}}
\newcommand{\efl}{\end{flushleft}}
\newcommand{\efr}{\end{flushright}}
\newcommand{\bc}{\begin{center}}
\newcommand{\ec}{\end{center}}
\newcommand{\be}{\begin{equation}}
\newcommand{\ee}{\end{equation}}
\newcommand{\bea}{\begin{eqnarray}}
\newcommand{\eea}{\end{eqnarray}}
\newcommand{\ba}{\begin{array}}
\newcommand{\ea}{\end{array}}
\newcommand{\nn}{\nonumber}
\newcommand{\edc}{\end{document}}
\newcommand{\ul}{\underline}
\newcommand{\ri}{\rightarrow\infty}
\newcommand{\li}{\leftarrow\infty}
\newcommand{\ra}{\rightarrow}
\newcommand{\la}{\leftarrow}
\newcommand{\ds}{\displaystyle}
\newcommand{\dsf}{\displaystyle\frac}
\newcommand{\dt}{\Delta{t}}
\newcommand{\il}{\int\limits}
\newcommand{\pal}{\partial}
\newcommand{\xxx}{{\it{X}}}
\newcommand{\bone}{{\bf 1}}
\newcommand{\gComment}[1]{}
\renewcommand{\gComment}[1]{\textcolor{red}{Gerardo: #1}}

\title[]{
Nanoscale Phenomenology from Visualizing Pair Formation Experiment}


\author{B. \surname{Abdullaev$^{1,2}$}}
\author{C. -H. \surname{Park$^2$}}
\author{M. M. \surname{Musakhanov$^{1}$}}

\affiliation{
 Physics, National University of
Uzbekistan, Tashkent 100174, Uzbekistan.\\
$^2$Research Center for Dielectric and Advanced Matter Physics,
Department of Physics,\\ Pusan National University, 30
Jangjeon-dong, Geumjeong-gu, Busan 609-735, Korea.}

\date{Received \today }

\begin{abstract}
Recently, Gomes {\it et} {\it al.}~\ci{Gomes} have visualized the
gap formation in nanoscale regions (NRs) above the critical
temperature $T_c$ in the high-$T_c$ superconductor
$Bi_2Sr_2CaCu_2O_{8+\delta}$. It has been found that, as
the temperature lowers, the NRs expand in the bulk superconducting state
consisted of inhomogeneities. The fact that the size of the
inhomogeneity~\ci{Pan} is close to the minimal size of the
NR~\ci{Gomes} leads to a conclusion that the superconducting phase
is a result of these overlapped NRs. In the present paper we
perform the charge and percolation regime analysis of NRs and show
that at the first critical doping $x_{c1}$, when the
superconductivity starts on, each NR carries the positive electric
charge one in units of electron charge, thus we attribute the NR
to a single hole boson, and the percolation lines connecting these
bosons emerge. At the second critical doping $x_{c2}$, when the
superconductivity disappears, our analysis demonstrates that the
charge of each NR equals two. The origin of $x_{c2}$ can be
understood by introducing additional normal phase hole fermions in
NRs, whose concentration appearing above $x_{c1}$ increases
smoothly with the doping and breaks the percolation lines of
bosons at $x_{c2}$. The last one results in disappearing the bulk
bosonic property of the pseudogap (PG) region, which explains the
upper bound for existence of vortices in Nernst effect~\ci{Wang}.
Since~\ci{Gomes} has demonstrated the absence of NRs at the PG
boundary one can conclude that along this boundary, as well as in
$x_{c2}$, all bosons disappear. 
\end{abstract}


\maketitle

\newpage

\section{Introduction}
\label{sec1}

The origin of PG and high-temperature superconductivity phases in
copper oxides is the most puzzling and challenging problem in
condensed matter physics. Despite on the intensive experimental
and theoretical studies we have no clear understanding of these
phases so far. A relationship between two phases has become a
subject of wide range theoretical proposals and their possible
experimental testing. A precursor scenario for the PG state
supposes pairing correlations without superconducting phase
coherence \ci{Emery}. This scenario has been confirmed by
experiments~\ci{Wang,Corson}. A description for the PG phase based
on the electronic competing order mechanism with experimental
arguments was given in~\ci{Tallon}. Other observations have
associated the PG with a real space electronic
organization~\ci{Vershinin} which is dominant at low dopings.

The fundamental property of the PG is a partial gap in the density
of states~\ci{Timusk} which is observed in various experiments.
To understand the nature of this gap the real space atomic scale
scanning tunneling microscopy measurements of the copper oxide
$Bi_2Sr_2CaCu_2O_{8+\delta}$ have been performed. For the case of high-$T_c$
superconductivity the spatial gap inhomogeneities have been observed
in~\ci{Howald,McElroy}, while Pan {\it et} {\it al.}~\ci{Pan}
explicitly determine their minimal size. The evolution of the
nanoscale gap formation with temperature decrease in the PG region
has been investigated by Gomes {\it et} {\it al.}~\ci{Gomes}.

In the present paper we study the origin of minimal size NRs,
which were visualized in Refs.~\ci{Gomes,Pan} through the
measurement of the energy gap. We use the experimental fact that
PG and superconductivity phases are formed from the NRs.
Particularly, we are interested in the electric charge of NRs. We
will employ the information about the charge to understand some
ingredients of doping-temperature phase diagram of
$Bi_2Sr_2CaCu_2O_{8+\delta}$ copper oxide. The generalization of
our consideration to other cuprates will be given as well. It is
worth to notice that all physical findings in the paper are
inferred from the analysis of data for the NRs in~\ci{Gomes,Pan}.
The most important fermionic nature of the second hole inside NR
at $x_{c2}$ and dopings below $x_{c2}$ is implied from the meaning
of the second critical doping $x_{c2}$: at this doping the
superconductivity and hence, the bosonic property of the matter
disappears. 

The authors of Ref.~\ci{Gomes} have visualized the NRs in the PG
region of $Bi_2Sr_2CaCu_2O_{8+\delta}$ compound at fixed hole
dopings $x=0.12,0.14,0.16,0.19,0.22$. It has been determined that
for $x=0.16$ and $x=0.22$ the minimal size of the NRs is
$\xi_{coh}\approx 1-3$ nm. The estimated minimal size of NRs,
$\xi_{coh}$, is about 1.3 nm in the superconducting phase~\ci{Pan}
($T_c=84 K$). Another notable result obtained in Ref. \ci{Pan} is
the observation of spatial localization of the dopped charges. The
charges are localized in the same area as NRs \ci{Pan} with the
same coherence length $\xi_{coh}$. Below we will demonstrate that
the next spatial parameter, the mean distance between two holes,
$r_0$, is important to understand the underlying physics. The
experimental doping dependence of $r_0$ can be approximated by the
relationship $r_0\approx a/x^{1/2}$ (see Fig. 34 in
Ref.~\ci{Kastner}), where $a$ is a lattice constant in the
elementary structural plaquette for the $CuO_2$ $a - b$ plane of a
copper oxide. This relationship is derived in Ref.~\ci{Kastner}
for $La_{2-x}Sr_xCuO_4$ compound with $a\approx 3.8 \AA$. It
is valid for our compound as well since the lattice constant $a$
of $Bi_2Sr_2CaCu_2O_{8+\delta}$ is $a\approx 3.8 \AA$ (see the
capture to Fig.2 in Ref.~\ci{McElroy}). It is worth to mention
that $b\approx a$ for the lattice constant $b$
of the same structural plaquette.

A principal part of our analysis is the doping $x$ dependence of
the NR charge $(\xi_{coh}/r_0)^2$. We start with
a case of zero temperature. The parameter
$\xi_{coh}/r_0$ contains an essential information in our
consideration. The factor $(\xi_{coh}/r_0)^2$ reduces to the
expression $x(\xi_{coh}/a)2$ which has a simple physical meaning:
it is a total electric charge of $(\xi_{coh}/a)^2$ number of
plaquettes each of them having a charge $x$. On the other hand,
the parameter $\xi_{coh}/r_0$ describes the average spatial
overlapping degree of two or more holes by one NR. If
$\xi_{coh}/r_0>1$ then all NRs will be in close contact to each
other providing by this the bulk superconductivity in percolation
regime.

In the Table I we outline the doping $x$ dependencies for the
function $(\xi_{coh}/r_0)^2$ for fixed experimental values
$\xi_{coh}=10\AA$ (the minimal size of the NR) and
$\xi_{coh}\approx13\AA$ taken from Ref.~\ci{Gomes} and
Ref.~\ci{Pan}, respectively, and for the function $\xi_{coh}$
which fits $(\xi_{coh}/r_0)^2$ to $(10\AA/r_0)^2$ at
$x=0.28$ and for $x=0.05$ provides $(\xi_{coh}/r_0)^2\approx 1.0$.
Numerical values of the $\xi_{coh}/r_0$ are also shown in the table.

Since in Ref.~\ci{Gomes} every NR location in the sample is
tracked with the precision $0.1\AA$, we suppose that the
$\xi_{coh}=10\AA$ has been measured with a high enough accuracy. In
addition, we assume that in Ref.~\ci{Pan} the $\xi_{coh}\approx
13\AA$ is measured with the same accuracy at $x=0.14$. Under this
condition, we conclude that the tendency of $\xi_{coh}$ to growth
from $10\AA$ to $13\AA$ when $x$ decreases from $0.22$ to $0.14$
reflects quantitatively the underlying physics. The data for
the resulting parameter $\xi_{coh}/r_0$ is approximated by the
function $2.2x^{1/3}$. The analytic equation for
$\xi_{coh}$ expressed in terms of the lattice constant $a$ is
given by $\xi_{coh}\approx 2.2 a/x^{1/6}$.

As seen from Table I, the charges $(10\AA/r_0)^2$,
$(13\AA/r_0)^2$, and $(\xi_{coh}/r_0)^2$ vary continuously with
the doping $x$. This is not surprising because they are functions
of $r_0(x)$ and $\xi_{coh}(x)$. From the analysis at
the first critical doping, $x_{c1}=0.05$, it follows that the
charge $(\xi_{coh}/r_0)^2$ of the visualized NR in Ref.~\ci{Gomes}
equals $+1$. So that, it corresponds to the charge of a single
hole. Notice, at the critical doping $x_{c1}=0.05$ the percolation parameter
is given by $\xi_{coh}/r_0=1.0$. That means the whole sample is entirely
covered with mini areas $\xi_{coh}^2=r_0^2$ contacting each other.
It is unexpected that at the second critical doping,
$x_{c2}=0.28$, the charge of the visualized NR takes the value
$+2$. This implies that at $\xi_{coh}^2=2r_0^2$ one has a pair of
holes inside the NR and, as a result, the superconductivity
disappears completely. For $x_{c2}=0.28$ we have
$\xi_{coh}/r_0>1.0$, so that the charge conductivity of the
fermions still remains.

Notice, that there are no particles
in the nature with the fractional charge, except the
quasiparticles which can be produced by many-body correlations
like in the fractional quantum Hall effect~\ci{Laughlin}. Hence,
the problem of the presence of the extra fractional charge inside
the NR has to be solved yet.
We remind~\ci{Gomes,Pan} that PG visualized NRs
constitute the bulk superconductivity phase below the critical
temperature $T_c$, and therefore, they are a precursor for that
phase. This implies undoubtedly that the NRs represent bosons at
least. At $x_{c1}=0.05$ one has the charge $(\xi_{coh}/r_0)^2=1$,
so that one may conjecture that the NR represents just a boson
localized in the square box $\xi_{coh}^2$.

For $x>0.05$ the charge $(\xi_{coh}/r_0)^2$ has an additional to
$+1$ fractional part. We assign the last one to the fractional
part of the charge of fermion. Thus the total charge
$(\xi_{coh}/r_0)^2$ of the NR includes the charge $+1$ of the
boson and the fractional charge of the fermion. However, as it was
mentioned above, the fractional charge can not exist. Therefore,
we take the number $N_{ob}$ of NRs to be equaled to the inverse
value of the fractional part to form a charge $+1$ of the fermion.
As a result, we obtain one fermion surrounding by $N_{ob}$ bosons.
The values of $N_{ob}$ are outlined in the last column of the
Table 1.

The NRs introduced in such a manner allow to understand clearly
the evolution of the fermions in the whole range $0.05 \leq x \leq
0.28$ of doping and to explain the origin of the second critical
doping $x_{c2}=0.28$. It is clear, as $x$ increases, the number of
fermions grows up inside the superconducting phase. By this, at
$x_{c2}$, when the number of fermions becomes equal to the number
of bosons, one has the breaking of the boson percolation lines,
and, thus the superconductivity disappears.

It is worthwhile to compare $\xi_{coh}$ with the lattice constant
$a$ of $Bi_2Sr_2CaCu_2O_{8+\delta}$ compound when the doping $x$
varies. We have $2.6a\leq \xi_{coh}\leq 4.5a$ for variation of $x$
from $x_{c2}$ to $x_{c1}$. However, it is well known that the
antiferromagnetic dielectric parent materials are characterized by
a strong short range magnetic interaction within the atomic length
scale $a$. Therefore, one may assume that $a$ is a length
parameter for these compounds. The fact that the size $\xi_{coh}$
is larger than $2.6a$  leads to a conclusion that the visualized
NRs are independent from the dielectric environment (the latter
forms only the spatial square shape of the NR). Due to this, the
numerical values $x_{c1}=0.05$ and $x_{c2}=0.28$ are universal for
all hole doped cuprates. However, at the second critical doping
$x_{c2}$ the length scale of boson and fermion (the half of
$\xi_{coh}$) inside NR is comparable to $a$. Therefore, the
parent compound starts to play a role from the critical doping
$x_{c2}$. Furthermore, since a coincidence of $x_{c2}$ with the PG
boundary at a zero temperature has been observed in various
experiments and for all temperatures of this boundary no NRs,
which exhibit gaps, were detected~\ci{Gomes}, the plausible
intuitive finding would be the total disappearance of bosons along
the PG bound line. So that two fundamental phenomena -- the
breaking of the boson percolation lines and the disappearance of
bosons -- occur at $x_{c2}$. The first phenomenon indicates the
end of the bulk bosonic property and the end of the $T_c$ curve as
well, whereas the second phenomenon corresponds to the end of the
bosonic property in general.  For the PG region the disappearance
of the bulk bosonic property was detected by observing the onset
temperature, $T_{onset}$, for the existence of vortices in the
Nernst effect~\ci{Wang}. The vortices have been seen so far only
in quantum Bose systems. Further evolution of fluctuations with
temperature increase destroys the bosons which totally vanish at
PG boundary.

\begin{table}[tb]
\begin {center}
\begin{tabular}{|c|c|c|c|c|c|c|} \hline
    $x$   & $(10 \AA/r_0)^2$ & $(13 \AA/r_0)^2$ & $\xi_{coh}(\AA)$ & $(\xi_{coh}/r_0)^2$ &
$\xi_{coh}/r_0$   & $N_{ob}$ \\ \hline
    0.28  &    1.939         &     3.277        &      10          &   1.939
& 1.393 &  $\sim 1$ \\ \hline
    0.22      &  1.524       &     2.575        &      10         &    1.524
& 1.235 &  $\sim 2$ \\ \hline
    0.16      &  1.108       &     1.873        &      11         &    1.341
& 1.158 &  $\sim 3$ \\ \hline
  0.14      &    0.969       &     1.638        &      12         &    1.396
& 1.182 &  $\sim 3$ \\ \hline
  0.10      &    0.693       &     1.170        &      13         &    1.170
& 1.082 &  $\sim 6$ \\ \hline
  0.05      &    0.346       &     0.585       &       17         &    1.000
& 1.000 &  \\ \hline
  0.04      &    0.277       &     0.468       &       18         &    0.897
& 0.947 &  \\ \hline
  0.02      &    0.139       &     0.234       &       20         &    0.554
& 0.744 &  \\ \hline
\end{tabular}
\end{center}
\vskip -.5cm \caption{The doping $x$ dependencies of NR charges. The
doping $x$ dependencies for $(10 \AA/r_0)^2$, $(13 \AA/r_0)^2$ at
fixed $\xi_{coh}=10 \AA$ and $\xi_{coh}=13 \AA$, respectively, for
the coherent length $\xi_{coh}$, the charge $(\xi_{coh}/r_0)^2$
and the percolation parameter $\xi_{coh}/r_0$ at this $\xi_{coh}$
are presented. The values for the number $N_{ob}$ of bosons
surrounding every fermion are shown in the last column.} \vskip
-.5cm \lab{tab-1}
\end{table}

\begin{figure}
\begin{center}
\includegraphics[width=8cm,scale=1]{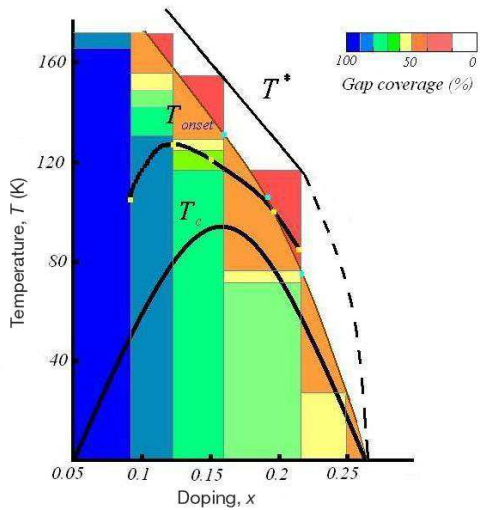}
\end{center}
\caption{Schematic single hole bosonic phase diagram for $Bi_2Sr_2CaCu_2O_{8+\delta}$.} \lab{fig1}
\end{figure}

The schematic single hole bosonic phase diagram for
$Bi_2Sr_2CaCu_2O_{8+\delta}$ is depicted in the Fig. 1. The
coloured zones indicate the percentage of the sample that is
gapped at given temperature and doping (in analogy with the phase
diagram shown in Ref.~\ci{Gomes}). The solid lines correspond
to the following observed temperatures: PG boundary
$T^*$ and onset temperature $T_{onset}$ for Nernst effect signals
taken from Ref.~\ci{Wang}, and the critical temperature $T_c$ from
Ref.~\ci{Gomes}. The extrapolation of the connection of $T^*$ with
the second critical doping, $x_{c2}$, is depicted by the dashed
line. The yellow points correspond to fixed $T_{onset}$ values
from Ref.~\ci{Wang}, and the blue points represent the temperature
data for $50\%$ of gapped area of the sample from Ref.~\ci{Gomes}
measured at fixed dopings. The thin brown coloured solid line fits
the blue points. The percentage for the gapped doping is
calculated by using the equation $(1-1/ (N_{ob}+1))\cdot 100\%$
under the assumption that the NRs overlap each other. It is
remarkable that $T_{onset}$ line is substantially located in the
brown coloured zones which means there is no bulk bosonic property
above these zones. It is worth to compare the homogeneous $100\%$
gap coverage observed in Ref.~\ci{Gomes} with our proposed varying
one in Fig. 1 for low temperature and doping levels $0.12\leq
x\leq 0.22$. Employing the doping changing dynamics of the $50\%$
gap coverage obtained in Ref.~\ci{Gomes}, we find that this
percentage is applicable also at the second critical doping,
$x_{c2}$, which is shown in Fig. 1 by yellow ($60\%$) and brown
($50\%$) colours. On the other hand, if we consider the NR charge
$+1$ for the above interval of doping with further its increasing
up to $+2$, close to $x_{c2}$, then we will reproduce exactly the
percentages observed in the phase diagram in Ref.~\ci{Gomes}.

The next interesting finding is that the number of external
interstitial atoms sufficient to produce one doped hole in the
dielectric parent material equals to $1/x$. For
$La_{2-x}Sr_xCuO_4$ compound it is a number of $Sr$ atoms, since
the hole doping and the concentration of atoms are expressed by
$x$. In the interval $0.05\leq x\leq 0.28$ this number varies from
$20$ to $3$.

We discuss on the percolation
threshold of $2D$ classical systems and compare it with our $50\%$
one used for bulk bosonic property. If we remind white and black
cells of the chessboard and assume that the black ones represent a
region in which the percolation should occur, it becomes clear
that the percolation threshold consists of $50\%$ coverage by this
colored region of the whole chessboard area. The experiment for
some particular system indicates to more than $40\%$ coverage for
its value~\ci{Smith} (the numerical simulation for the same system
has confirmed this observed result~\ci{Weinrib}).

Summarizing the paper, we have succeeded in understanding the
following constituents of the doping-temperature phase diagram of
the hole doped copper oxides: (i) the first and second critical
dopings have been a result of emergence and disappearance of the
single hole boson percolation lines, respectively; (ii) the
disappearance of the percolation lines leads to the end of the PG
bulk bosonic property or to the end of Nernst effect signals;
(iii) the fact that the PG boundary was a bound, where the single
hole bosons disappear, confirmed by Ref.~\ci{Gomes}. Our findings
are consistent with the recent observation~\ci{Gavrilkin} of the
superconducting phase consisted of the array of nanoclusters
embedded in the insulating matrix and of percolative transition to
this phase from the normal phase in $YBa_2Cu_3O_{6+ \delta}$.
Superconducting islands introduced in insulating background
have been used for the interpretation of the
superconductor-insulator transition in
$Bi_2Sr_{2-x}La_xCaCu_2O_{8+\delta}$ compound~\ci{Oh}. In a recent
paper~\ci{tahir} a significant role of
the percolation of elementary structural plaquettes on universal
properties of cuprates has been established. Using $3D$
percolation mechanism the authors of Ref.~\ci{tahir} succeeded in
explanation of the  $T_c$ phase diagram, room-temperature
thermopower, neutron spin resonance, and STM incommensurability.
At the end, the boson and fermion mixing
nature of PG region, derived from experiment~\ci{Gomes}, is
consistent with our treatment and description of low temperature
non-Fermi liquid heat conductivity and entropy~\ci{Abdullaev2}.

\section{Acknowledgements}

The work is partially supported by Korean Research Foundation
(Grant KRF-2006-005-J02804).


\begin{thebibliography}{0}

\bibitem{Gomes}
K. K. Gomes {\it et} {\it al.}, Nature 447 (2007) 569.

\bibitem{Pan}
S. H. Pan {\it et} {\it al.}, Nature 413 (2001) 282.

\bibitem{Wang}
Y. Wang, L. Li, and N. P. Ong, Phys. Rev. B 73 (2006) 024510.

\bibitem{Emery}
V. J. Emery and S. A. Kivelson, Nature 374 (1995) 434.

\bibitem{Corson}
J. Corson {\it et} {\it al.}, Nature 398 (1999) 221; Z. A. Xu {\it
et} {\it al.}, Nature 406 (2000) 486; Y. Wang {\it et} {\it al.},
Phys. Rev. Lett. 95 (2005) 247002.

\bibitem{Tallon}
J. L. Tallon and J. W. Loram, Physica C 349 (2001) 53.

\bibitem{Vershinin}
M. Vershinin {\it et} {\it al.}, Science 303 (2004) 1995; T.
Hanaguri {\it et} {\it al.}, Nature 430 (2004) 1001; K. McElroy {\it
et} {\it al.}, Phys. Rev. Lett. 94 (2005) 197005.



\bibitem{Timusk}
N. Timusk and B. Statt, Rep. Prog. Phys. 62 (1999) 61.

\bibitem{Howald}
C. Howald, P. Fournier, and A. Kapitulnik, Phys. Rev. B 64 (2001)
100504(R).

\bibitem{McElroy}
K. McElroy {\it et} {\it al.}, Science 309 (2005) 1048.


\bibitem{Kastner}
M. A. Kastner, R. J. Birgeneau, G. Shirane, and Y. Endoh, Rev. Mod.
Phys. 70 (1998) 897.

\bibitem{Laughlin}
R. B. Laughlin, in {\it The Quantum Hall Effect,} Edited by R. E.
Prange and S. M. Girvin, (Springer-Verlag, New York, 1987).

\bibitem{Smith}
L. N. Smith and C. J. Lobb, Phys. Rev. B 20 (1979) 3653.

\bibitem{Weinrib}
A. Weinrib,  Phys. Rev. B 26 (1982)1352.


\bibitem{Gavrilkin}
S. Yu. Gavrilkin, O. M. Ivanenko, V. P. Martovitskii, K. V.
Mitsen, and A. Yu. Tsvetkov, Arxiv: 0909.0612.

\bibitem{Oh}
S. Oh, T. A. Crane, D. J. Van Harlingen, and J. N. Eckstein, Phys.
Rev. Lett. 96 (2006) 107003.

\bibitem{tahir}
J. Tahir-Kheli and W. A. Goddard, J. Phys. Chem. Lett. 1 (2010)
1290.

\bibitem{Abdullaev2}
B. Abdullaev, C. -H. Park, and K. -S. Park, cond-mat/0703290.

\end{thebibliography}
\end{document}